# Three-dimensional light capsule enclosing perfect super-sized darkness via anti-resolution

Chao Wan[1]*, Kun Huang[1]*, Tiancheng Han[1]*, Eunice S. P. Leong[2], Weiqiang Ding[1], Tat-Soon Yeo[1], Xia Yu[3], Jinghua Teng[2], Dang Yuan Lei[4], Stefan A. Maier[5], Shuang Zhang[6] and Cheng-Wei Qiu[1]†

[1]Department of Electrical and Computer Engineering, National University of Singapore, 4 Engineering Drive 3, Singapore 117576, Singapore
[2]Institute of Materials Research and Engineering, Agency for Science, Technology and Research, 3 Research Link, Singapore 117602, Singapore
[3]Singapore Institute of Manufacturing Technology, Agency for Science, Technology and Research, 71 Nanyang Drive, Singapore 638075, Singapore
[4]Department of Applied Physics, The Hong Kong Polytechnic University, Kowloon, Hong Kong
[5]The Blackett Laboratory, Department of Physics, Imperial College London, London SW7 2 AZ, UK
[6]School of Physics & Astronomy, University of Birmingham, Birmingham, B15 2TT, UK

We theoretically and experimentally demonstrate the focusing of macroscopic 3D darkness surrounded by all light in free space. The object staying in the darkness is similar to staying in an empty light capsule because light just bypasses it by resorting to destructive interference. Its functionality of controlling the direction of energy flux of light macroscopically is fascinating, similar in some sense to the transformation-based cloaking effect. Binary-optical system exhibiting anti-resolution (AR) is designed and fabricated, by which electromagnetic energy flux avoids and bends smoothly around a nearly perfect darkness region. AR remains an unexplored topic hitherto, in contrast to the super-resolution for realizing high spatial resolution. This novel scheme replies on smearing out the PSF and thus poses less stringent limitations upon the object's size and position since the created dark (zero-field) area reach 8 orders of magnitude larger than $\lambda^2$ in size. It functions very well at arbitrarily polarized beams in three dimensions, which is also frequency-scalable in the whole electromagnetic spectrum.

It is well known that the point spreading function (PSF) (Fig. 1a) dictates the performance of an optic focusing system [1,2], so a sharp PSF with strong main lobe and weak sidelobe (Fig. 1b) is highly desired for realizing high spatial resolution [3]. In the past decade, much research effort has been dedicated to narrowing the PSF so as to achieve super-resolution and beat the diffraction limit [4-13]. Concurrently, significant effort in optical super-resolution has been devoted to the development of various optical microscopy techniques (e.g., stimulated emission depletion microscopy [12] and stochastic optical reconstruction microscopy [13]) based on molecular labeling, nonlinear optical saturation, luminescence, and excitation/de-excitation of fluorophores.

One interesting question beyond the current focus is what new interesting phenomena and applications exist if one pushes the limit toward the other extreme, i.e., suppressing and flattening the main lobe in the PSF until it has completely vanished and elevating the sidelobe (Fig. 1d), which is completely inverse to the manipulation of super-resolution in Fig. 1b. For this anti-resolution PSF, a point source at the object plane leads to a large-radius ring intensity, resulting in the dis-resolved imaging at the imaging plane. This new scheme, defined in our concept as anti-resolution (AR), can be obtained by designing an imaging system made of concentric dielectric grooves together with a focusing lens. Such implementation creates a macroscopic spatial region with nearly perfect dark region where the energy flux of light approaches zero. The darkness in PSF spreads in the focal region for the uniform variation of energy flux in the homogenous medium, resulting in a huge darkness region surrounded by light, which looks like three-dimensional "optical capsule". It enables the object inside to be dis-resolved or undetectable in this supersized darkness region of optical capsule. A three-dimensional object placed in the optical capsule does not cause scattering and one can therefore see the scene behind the object. The challenge is two-fold: (1) the dark region must have a remarkably large volume of smeared-out PSF; (2) the light intensity inside this volume must be thoroughly diminished.

Here we theoretically and experimentally demonstrate that such a huge three-dimensional light capsule is not a fantasy, but that it can be created using a 3D binary-optical system composed of dielectric grooves in conjunction with a focusing lens, through suppressing the main spot to invert it upside down in PSF and widening the main spot as much as possible – a concept of anti-resolution. It is important to note that the 3D distribution of light is re-arranged in air without superluminal propagation [14], and exhibits the feature of self-imaging beams in a physical configuration that is distinguished from the classical mathematics-based Guoy effect [15]. It is further shown in our experiment that the focused darkness with nearly zero-field intensity, embraced by surrounding light, can be $1.6 \times 10^{10} \lambda^2$ (with λ being the wavelength of the input light). This unique scheme of light capsule is free of demanding nanofabrication [16], and more importantly, the center position of the darkness can robustly controlled simply by numerical aperture of the involved focusing lens.

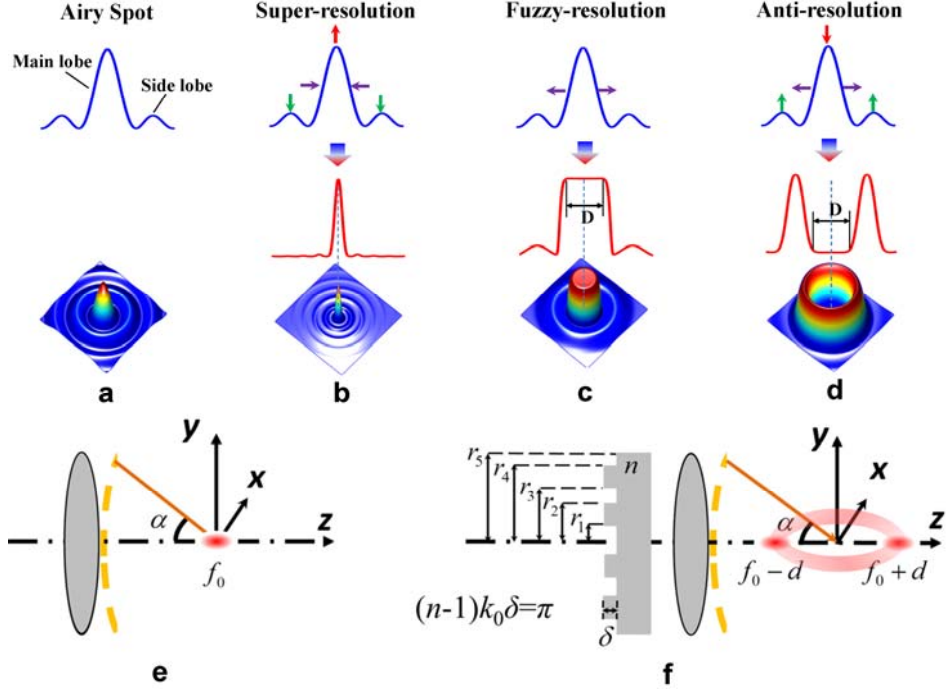

FIG. 1. Evolution from super-resolution to anti-resolution. The PSF of a traditional optical imaging system has the form of Airy spot (a). The super-resolution in (b) is achieved by narrowing the PSF, enhancing the mainlobe as well as suppressing the sidelobe. If we just widen the PSF without any disposing of mainlobe and sidelobe in PSF, the fuzzy-resolution (c) leading to degradation in imaging can be obtained. One completely inverse case of (b) in manipulation of PSF is widening the PSF, suppressing the mainlobe and enhancing the sidelobe so that the PSF in the region D ($\gg \lambda$) completely vanishes, which is the concept of anti-resolution (d). (e) Focusing by a conventional lens with binary phase off. (f) The scheme for realizing optical capsule based on anti-resolution with binary phase on.

In optics, the interconnection between the input pattern $O(x,y)$ in the object space and the output pattern $I(x,y)$ in the image space is described with the help of a point spread function (PSF) $P(x,y)$ according to the following relation

$$I(x,y) = \iint_{\text{object plane}} O(\xi,\eta) P(x-\xi, y-\eta) d\xi d\eta \tag{1}$$

Conventional optical imaging is usually concerned with the PSF's width approaching zero in order to beat the diffraction limit corresponding to Fig. 1b. One interesting question beyond current research focus (engineering Fig. 1b and 1c) is what new phenomena and applications exist if one pushes the limit toward the other extreme, i.e., broadening the PSF as in Fig. 1d, suppressing the mainlobe until it vanishes completely and enhancing the sidelobe as shown in

Fig. 1d. Such a scheme, defined in our concept as anti-resolution (AR), can convert "positive" light distribution into "negative" counterpart - darkness, i.e. as a three-dimensional empty light capsule. The fundamental relation between the PSF for super-resolution optics and light capsule is schematically presented in Fig. 1a-1d. Such a transformation of the PSF can be achieved by putting a 0-π phase lens with a special design of concentric rings, in front of another focusing lens, which works for incident lights of various polarization states, e.g., radially, azimuthally or linearly polarized beams. It is well known that the conventional focusing lens as in Fig. 1e will simply focus the light to the focal point ($f_0$), while Fig. 1f presents the physical scheme for realizing anti-resolution based light capsule, composed of multi-belt dielectric rings and a focusing lens. Based on the precise design of the binary ring belts, such dark region surrounded by the light can be formed around the original focal point.

Let us take a Bessel-Gauss beam, the focusing of which we describe via vector diffraction theory [17], as an example. The electric fields near the focus can be obtained for radially polarized light as:

$$E_r = A \int_0^\alpha \sqrt{\cos\theta} \sin(2\theta) \ell(\theta) T(\theta) J_1(kr\sin\theta) e^{ikz\cos\theta} d\theta \qquad (2)$$

$$E_z = 2iA \int_0^\alpha \sqrt{\cos\theta} \sin^2\theta \, \ell(\theta) T(\theta) J_0(kr\sin\theta) e^{ikz\cos\theta} d\theta \qquad (3)$$

or for azimuthally polarized light:

$$E_\varphi = 2A \int_0^\alpha \sqrt{\cos\theta} \sin\theta \, \ell(\theta) T(\theta) J_1(kr\sin\theta) e^{ikz\cos\theta} d\theta \qquad (4)$$

where $\ell(\theta) = \exp\left[-\beta^2\left(\frac{\sin\theta}{\sin\alpha}\right)^2\right] J_1\left(2\beta\frac{\sin\theta}{\sin\alpha}\right)$. Here $k = 2\pi/\lambda$ is the wave vector of light, $\alpha = \arcsin(NA)$, where NA denotes the numerical aperture of the focusing lens. $J_n(x)$ is the $n$-th order Bessel function of the first kind, and $\beta$ is the ratio of the pupil radius and the width of the beam waist. $A$ is a constant related to the focal length and the wavelength. $\ell(\theta) = 1$ represents uniform illumination, and $T(\theta) = e^{i\phi(\theta)}$ denotes the transmission function of the binary lens shown in Fig. 1f. A multi-belt groove is used as to modulate the phase of the incident light, with $\phi(\theta) = 0$ or $\phi(\theta) = \pi$ at corresponding ranges of angle $\theta$, i.e.,

$$T(\theta) = \begin{cases} 1, & \text{for } 0 < \theta < \theta_1, \theta_2 < \theta < \theta_3, \theta_4 < \theta < \alpha, \\ -1, & \text{for } \theta_1 < \theta < \theta_2, \theta_3 < \theta < \theta_4 \end{cases} \qquad (5)$$

The angles $\theta_i$ correspond to individual radii $r_i = \sin\theta_i/NA$ (normalized by radius $R$ of the lens). This type of phase modulation was used previously to create longitudinally polarized light [18]. Now we want to calculate a set of $\theta_i$ which will produce anti-resolution, resulting in the macroscopic perfect darkness where the destructive interference happens by using the 0-π phase modulation of binary lens. We find that the angles $\theta_i$ are the physical solution of a non-linear matrix equation that obeys the gauge of optical vectorial focusing in Eqs. (2-4) [19]. We can obtain the angles $\theta_i$ for realizing the anti-resolution, without any optimization, by solving the non-linear matrix equation numerically, which is well developed by using Newton theory [20]. It is worthy to point out that one set design of 5-belt binary-phase plate takes only about 2.8 seconds by solving the nonlinear matrix equation numerically on a 32G-RAM and 8-core-CPU (Intel Core i7) personal computer. The detailed process to solve the non-linear matrix equation is provided in Supplementary Materials. The field distributions for such a phase mask designed by the mentioned technique are shown in Fig. 2, which has a clear guideline to follow without resorting to blinded optimization. The envelope of the AR region (zero intensity) can be seen from the field cross sections. It is interesting to note that the electric field on the beam axis ($r = 0$) is longitudinally polarized similar to Ref. [18]. We can see also that the AR performance of the imaging system is independent of the polarization state of the incident light: it is practically the same for radially and azimuthally polarized light.

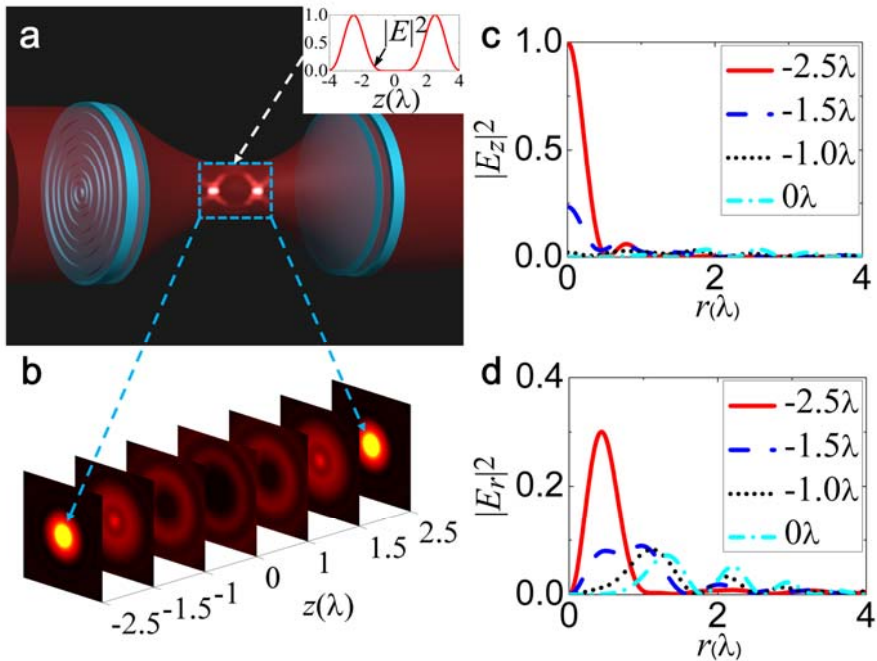

FIG. 2. Physical configuration and field distribution of the 3D light capsule. (a) Physical configuration for realizing light capsule, where a Bessel-Gauss beam propagates through a specially designed pairs of binary mask and focusing lens from the left to right (2 pairs). The inset shows the magnitude of the total electric field along the optical axis, in which a nearly perfect null region (the field amplitude at the order of $10^{-5}$) is formed. (b) 7 cut planes of equal separation distances within the dashed box in (a) are selected to demonstrate individual transversal field intensities. (c) Longitudinal polarized field intensity on the first 4 transversal cut planes. (d) Radially polarized field intensity on the first 4 transversal cut planes. The cut plane at z = 0 corresponds to the middle plane between the first focus (z = - 2.5 $\lambda_0$) and second focus (z = 2.5 $\lambda_0$).

Figure 2 presents the physical performance of 3D light capsule with a supersized darkness enclosed. The thickness of concentric dielectric grooves can be designed such that light passing through the groove has an additional π-phase difference compared to that passing through the neighboring air belts [18, 21]. A completely null region in the focal region is formed, i.e., the mainlobe in the PSF vanishes in this region. As a result, the incident probe light has no interaction with the object placed within the null region and just travels around the object. The same lens and complementary $SiO_2$ grooves can be positioned symmetrically on the other side of the focus point, restoring the incident beam's wavefront. Note that the same system has been validated for a variety of polarizations and wavefronts, e.g., plane waves and Bessel-Gauss beams of different polarization.

Figure 2b presents the field distributions on seven cross-sectional planes at specified locations from the left to the right imaging planes. The envelope of the AR region (the null field) can be seen from the varying sizes of smeared-out regions in Fig. 2b. Note that the electric field on the beam axis (*r* = 0) is purely longitudinally polarized ($E_z$ is the only electric-field component) as Fig. 2c shows. Since such polarization is parallel to the propagation direction, at first glance it seems to violate Maxwell's equations governing the propagation. Nevertheless, when the position is deviating from the beam axis (*r* ≠ 0), a radial component $E_r$ starts to emerge as indicated by Fig. 2d, which plays an important role in enabling the energy flux of light to bypass the dark region and reach the second focus. At *z* = -2.5$\lambda_0$ (i.e., the first imaging plane), $E_z$ clearly dominates within the area close to the beam axis, and therefore the energy of light is prohibited from flowing straightforwardly along the beam axis and has to travel in a curvilinear trajectory. At *z* = 0$\lambda_0$ (in the middle of two imaging planes), Fig. 2d shows that $E_r$ (corresponding to propagation) is almost smeared out within the central region around the beam axis. When it is more deviated from the center axis, $E_r$ gradually arises, and it corresponds to the propagation and the first color ring in the cut plane at $z = 0\lambda_0$ in Fig. 2b.

Simulations of the Poynting vector $\mathbf{S} = \dfrac{c}{4\pi}\text{Re}(\mathbf{E} \times \mathbf{H})$ field distribution in Fig. 3 verify the nearly negligible energy in the AR region (~10 orders less than that of the surrounding fields) when an object of the size of one–wavelength radius is placed in the AR region. For exactly the same binary lens and focusing lens as used in Fig. 2a, we consider radially Eq. (2) and azimuthally Eq. (3) polarized incident lights in Fig. 3a and Fig. 3b, respectively. The bifurcation and reformation of energy flux unambiguously reveals that the AR performance is nearly unaffected by the incident polarization states. Hence, any complex polarization state constituted vectorially with radial and azimuthal polarizations can all behave perfectly to yield identical SAR effect. It is also found that the size of AR region (zero-field darkness) can be extremely enlarged by reducing the NA value of the focusing lens (~$1/\text{NA}^3$), while preserving zero intensity in the dark focal field.

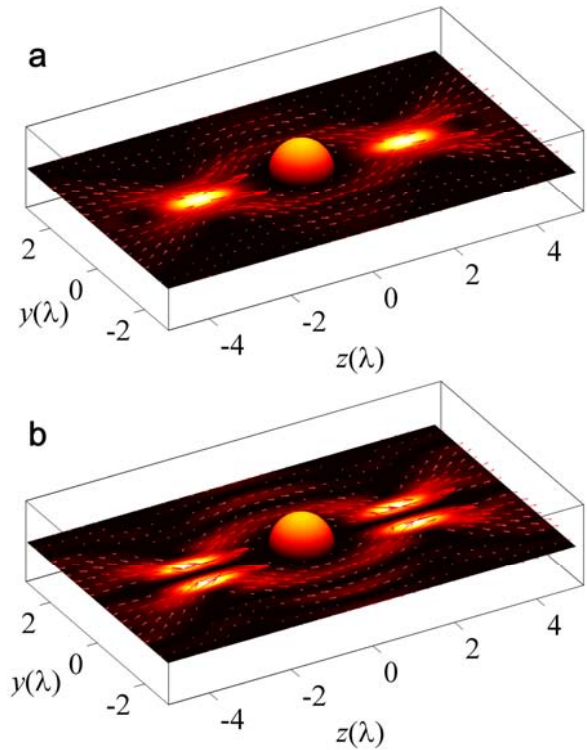

FIG. 3. Full-wave simulations for energy flux bifurcation and reformation in light capsule system at NA=0.95. (a) Incident light is radial polarized. (b) Incident light is azimuthally polarized. A metallic sphere with radius of one wavelength is placed in the center.

With the experimental setup schematically shown in Fig. 4a, we have experimentally demonstrated an AR-based light capsule embracing significantly large darkness. A pupil made of an opaque gold circular disk (150μm diameter) is placed at the center of the AR region, which blocks the straight line of sight and serves as an object to be detected. The photograph of fabricated binary mask is shown in Supplementary Fig. S1, and its phase profile is demonstrated in Fig. 4e. A gold plate with a letter 'N' etched through is placed behind the pupil to verify the reformation of light bypassing the pupil. The image of the letter 'N' captured by the CCD camera in Fig. 4b unambiguously evidences that the light can bend and bypass the larger opaque pupil. As a control experiment, when the binary lens is removed, light is fully blocked by the opaque disk of the pupil, resulting in very limited light reaching the letter of the object beneath, and consequently a very dark "N" as shown in Fig. 4c. The envelope of the darkness has been experimentally recorded in Fig. 4d.

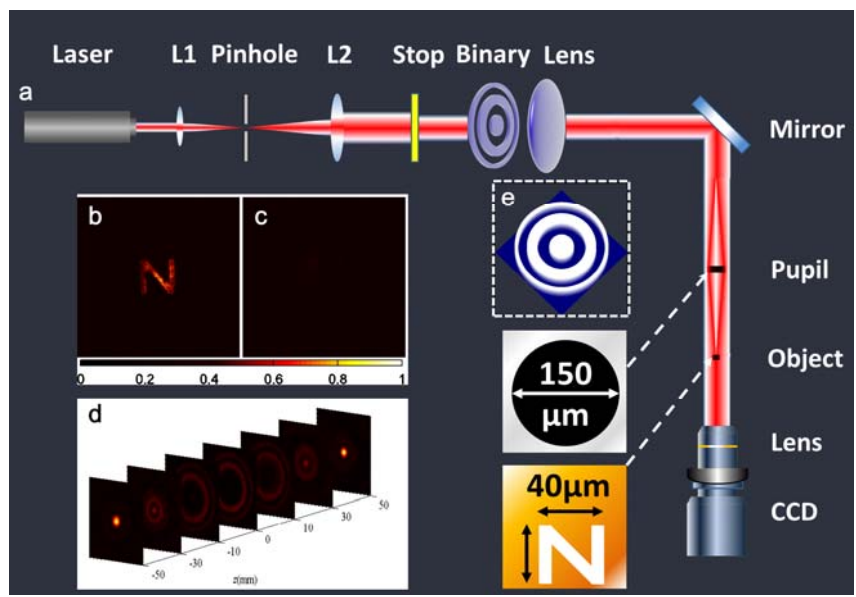

FIG 4. Experimental setup and measurement results for the SAR-based light capsule enclosing super-size darkness region in the imaging system. (a) Schematic of the experiment. Two insets represent the dimension of the opaque pupil and the size of the letter beneath the object, respectively. (b) When the mask phase and the lens phase are simultaneously adopted, light can bypass the opaque pupil and illuminate the object beneath. (c) When either mask or lens phase is adopted separately (i.e., lens or 0-π modulation mask is removed), the light is blocked by the opaque circular area (150μm in diameter). Almost no light passes the pupil and a very dark "N" is captured by CCD. A video (provided as a supplemental multimedia file) shows the switching of the bright and dark "N", corresponding to situations of the presence and absence of 0-π mask phase. (d) Measured intensity on 7 transversal planes, on both sides of the central plane ($z=0$). (e) The phase profile of fabricated binary mask shown in Fig. S1.

With the experimental setup unchanged, a spatial light modulator (SLM) is employed to represent the equivalent phase modulation exerted by the ring belts and lens, as shown in Supplementary Fig. S2a. The phase profile, equivalent to the ring belts and lens, is shown in

Fig. S2e. The measurement results are demonstrated in Figs. S2c and S2d, which are in good agreement with results using actual fabricated binary mask in Fig. 4.

Figure 5 shows four sets of designed parameters (Fig. 5a) by our optimization-free method and the size (the transversal D in Fig. 5b and axial $d_0$ in Fig. 5c) dependence of the optical capsule for every case on the NA of the focusing lens. Our theory predicts that the transversal (D) and axial ($d_0$) sizes of optical capsule are in proportion to 1/NA and 1/NA$^2$, respectively, which is shown in Fig. 5 and supplementary. Therefore, the size of the optical capsule can be gigantic for low NA. It implies that our design of AR-based 3D light capsule can give rise to a supersized darkness, still surrounded by the light. It unanimously demonstrates the importance of the binary $SiO_2$ grooves in the creation of nearly perfect darkness enclosed by all light surroundings, which can significantly manipulate the size of darkness embraced by the light while the central location ($z=f_0$) of the darkness and focusing lens' NA in Fig. 1f are unchanged. In Fig. 5b-5c, the fact that one binary-phase plate works very well for all the focusing lenses with different NA in generating the optical capsule, unveils the intrinsic property of the well-designed binary-phase plate: it can still generate the optical capsule with the huge-size darkness at its far-field (Fraunhofer) diffraction region even if the focusing lens does not exist in Fig. 1f. In fact, the role that the focusing lens in Fig. 1f has played is just pulling the far-field diffraction region of the binary-phase plate to the region near the focal plane of the focusing lens, which has been confirmed in Fourier optics [1]. As a result, the size of optical capsule near the focal region is tightly dependent on the lens's functionality: the weak (low NA) focusing lens leads to an optical capsule with large-size darkness while the strong (high NA) focusing lens results in an optical capsule with small-size darkness, which is proved by Fig. 5b-5c. Therefore, in physics, the most primitive root for the generation of optical capsule presented in the scheme of Fig. 1f is the binary-phase plate's substantial ability in forming the gigantic darkness (impossible to be detected directly without the focusing lens) surrounded by light at the far-field diffraction region.

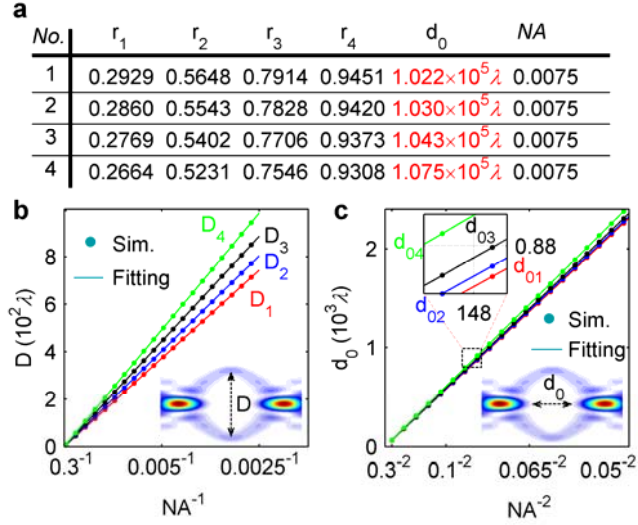

FIG 5. The ring belts design and its scaling properties of optical capsule for the lens with different NA. (a) Four sets of parameters that demonstrate the robust design of super-sized AR-based light capsule by using the optimization-free method. (b) The radial size (D) of null field in optical capsule generated by using the lens with different NA (from 0.3 to 0.0025) and the four sets of binary-phase plates in (a). Their fitting curves for different cases are: $D_1=1.8600\lambda/NA$ for No. 1 set of ring belt, $D_2=2.0117\lambda/NA$ for No. 2, $D_3=2.2171\lambda/NA$ for No. 3 and $D_4=2.4605\lambda/NA$ for No. 4. For all the four cases, the root-mean-square errors (RMSE) between the original data and fitting curves have the order of magnitude $10^{-14}$, indicating a perfect proportion of radial size D to $1/NA$. (c) The axial size ($d_0$) of null field in optical capsule generated by using the lens with different NA (from 0.3 to 0.05) and four sets of binary-phase plates in (a). Their corresponding fitting curves are: $d_{01}=5.7728\lambda/NA^2$, $d_{02}=5.8170\lambda/NA^2$, $d_{03}=5.8944\lambda/NA^2$ and $d_{04}=6.0762\lambda/NA^2$ with their fitting RMSEs at the order of magnitude $10^{-12}$, implying that the axial size $d_0$ is proportional to $1/NA^2$.

In summary, an optical scheme for realizing super-sized light capsule by creating an anti-resolution region with a binary phase (0-π) mask was presented theoretically and experimentally in air. The scheme is also adjustable, namely, the volume of darkness inside the light capsule can be changed in a drastically wide range, while the binary phase mask can be kept unchanged (only depending on NA of the focusing lens). We stress that the "bending of light" is in dramatic contrast to the Airy beam that is actively studied in plasmonics [22,23]. Our proposal of "light bending" is immune to energy loss, narrow operation bandwidth, limited SAR area, superluminal propagation, or polarization sensitivity. The AR scheme is also different from transformation-based cloaking schemes [16, 24, 25]. This new scheme of maneuvering light creates a plethora of possibilities for optical imaging systems, superb surveillance by seeing things behind for the military use, or cloaking the object surrounded by high field intensity.

C.W, K. H. and T.C.H. contributed equally to this work. This work is partly supported by Defence Science & Technology Agency (DSTA) in Republic of Singapore. C.W. acknowledges the financial support by CSC PhD scholarship. C.W.Q. acknowledges the Grant R-263-000-A23-232 administered by National University of Singapore. S.Z. and S.A.M. acknowledges the financial support by the Engineering and Physical Sciences Council of United Kingdom. We thank Dr. Hong Liu in IMRE for preparing the object and pupil, Dr. Haifeng Wang in DSI for discussion, and staffs in SIMTech for providing the facilities for measurement.

†eleqc@nus.edu.sg

# Supplementary Information

# Three-dimensional light capsule enclosing perfect super-sized darkness via anti-resolution

Chao Wan[1]*, Kun Huang[1]*, Tiancheng Han[1]*, Eunice S. P. Leong[2], Weiqiang Ding[1], Tat-Soon Yeo[1], Xia Yu[3], Jinghua Teng[2], Dang Yuan Lei[4], Stefan A. Maier[5], Shuang Zhang[6] and Cheng-Wei Qiu[1]†

[1]Department of Electrical and Computer Engineering, National University of Singapore, 4 Engineering Drive 3, Singapore 117576, Singapore
[2]Institute of Materials Research and Engineering, Agency for Science, Technology and Research, 3 Research Link, Singapore 117602, Singapore
[3]Singapore Institute of Manufacturing Technology, Agency for Science, Technology and Research, 71 Nanyang Drive, Singapore 638075, Singapore
[4]Department of Applied Physics, The Hong Kong Polytechnic University, Kowloon, Hong Kong
[5]The Blackett Laboratory, Department of Physics, Imperial College London, London SW7 2 AZ, UK
[6]School of Physics & Astronomy, University of Birmingham, Birmingham, B15 2TT, UK

## 1. Fabrication of the binary phase mask

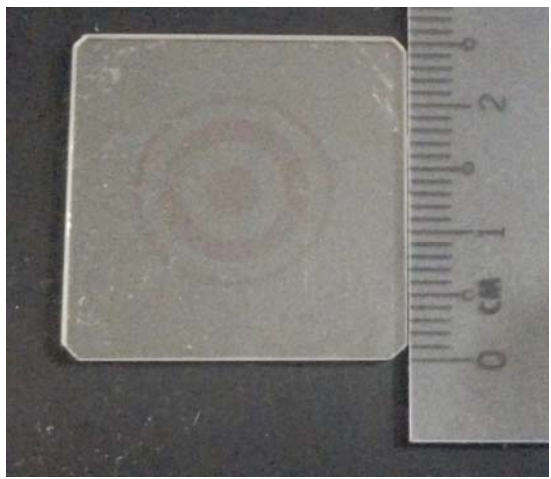

Figure S1. Photograph of the fabricated binary mask.

A 25.4×25.4×1mm quartz substrate is first cleaned with acetone and Isopropyl Alcohol (IPA) in an ultrasonic bath. AZ5214 resist is then spin coated on the substrate and pre-baked at 90°C for 30 minutes to remove the solvent. The sample was exposed under UV light in a bond aligner (SUSS MicroTec, MA8/BA6) with a photomask. The exposed regions were removed in AZ developer. After developing, the sample was post-baked at 120°C for 30 minutes. $MgF_2$ was evaporated onto the patterned sample using an e-beam evaporator (Denton Vacuum, Explorer). The base pressure was 1.1e-6 Torr and the thickness of the film was 826 nm. After evaporation, lift-off process was carried out in acetone to form the binary phase lens. The photograph of fabricated binary mask is shown in Fig. S1.

## 2. Reference experiment with spatial light modulator

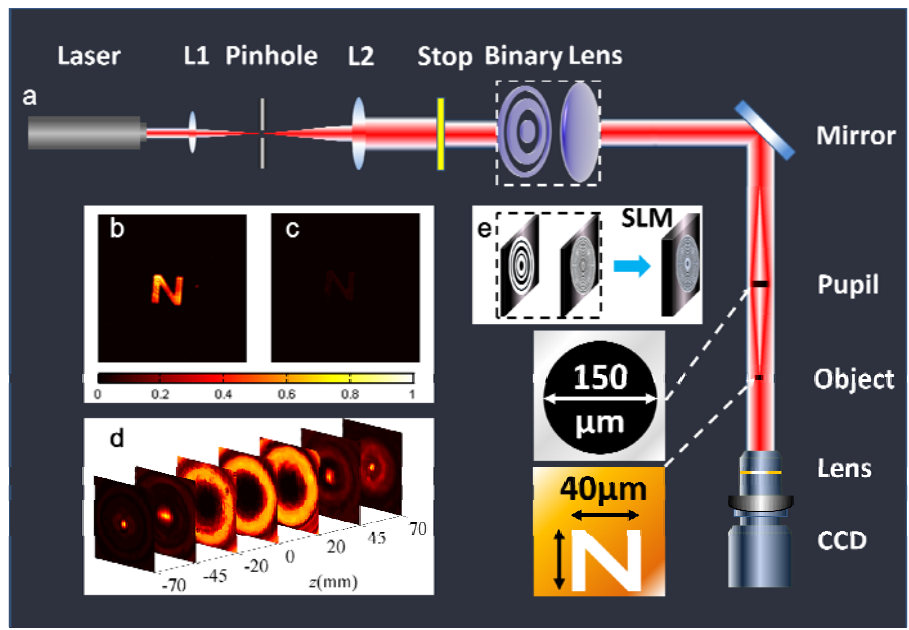

Figure S2. Reference experiment when the binary mask is replaced with a spatial light modulator. (a) Schematic of the experiment. (b) When the mask phase and the lens phase are simultaneously adopted by the SLM, light can bypass the opaque pupil and illuminate the object beneath. (c) When either mask or lens phase is adopted separately, the light is blocked by the opaque circular area. Almost no light passes the pupil and a very dark "N" is captured by CCD. (d) Measured intensity on 7 transversal planes, on both sides of the central plane ($z$=0). (e) The total phase profile of binary and lens is mimicked with SLM.

## 3. Experimental Section

The experiment is carried out on an optical table of 3 m long. A linearly polarized incident laser is expanded by a beam expander comprised of two lenses and one 10 μm-diameter pinhole. After going through a stop, the light illuminates the binary phase mask and a focusing lens (or equivalent SLM). A mirror is used to change the optical path considering the limitation of table length. An object (letter 'N') is located behind the pupil. The distance between the pupil and object is 20 mm. A CCD camera (IDS UI-2240) with a 20× camera lens is used to record the intensity.

In the experiment using binary phase mask fabricated on the quartz substrate, the wavelength of incident light is 632.8nm and the focusing lens has the NA of 0.0075 with its focal length of 1m and the size of 15mm in diameter, which is the same with the stop size and the maximum diameter of binary phase mask. The phase delay π between two neighboring belts in the mask is realized by their height difference of 692nm since the refraction index of quartz is 1.4507 at 632.8nm.

In the experiment using SLM, the phase modulation mask is created by a phase-type HOLOEYE LC2002 with a pixel pitch of 32 μm at the wavelength 532nm. A small gap with size of about 8 μm exists between two neighboring pixels. When the light passing through the gap is focused by a lens, a spot will be generated at the focal plane of focusing lens regardless of the phase of SLM. Since a low numerical-aperture (NA) lens is equivalent to a lens phase of $\exp[-i\pi r^2/(f\lambda)]$ in Fourier optics with $f$ being the focal length, the presence of the NA in Fig. 2a is thus replaced by a lens-phase to be equivalently realized by the SLM. The SLM produces the mask phase and the lens phase with $f = 2$m simultaneously. The stop has the size of 19.2mm so that the lens realized by SLM has the NA of 0.0048.

**4. Design of binary phase plate for realizing the anti-resolution PSF**

In the main text, Eqs (2-4) demonstrates the tight focusing of the vector beams (i.e. radially and azimuthally polarized beams) by using a high NA lens. In tightly focusing of vector beams, the electric field in the focal region has the strong dependence on the polarization of incident vector beams, e.g. Eqs. (2-3) for radially polarized beam and Eq. (4) for azimuthally polarized beam [1]. It is not convenient to describe the design of binary phase in a general way. For the sake of simplicity, we take the focusing lens with low NA for example. In this case, the polarization effect in the focal region is not significant and can be neglected so that, for the incident beams with various polarizations, the electric field in the focal region can be described in a unified way by using a scale focusing of light with a low

NA lens. For the focusing of a low NA circular lens with focal length $f$, the optical field in the focal region can be expressed by [2]

$$U(\rho,z) = \frac{i2\pi}{\lambda z} e^{ik(z+\frac{\rho^2}{2z})} \int_0^R u_0(r) \cdot e^{\frac{ikr^2}{2z}} e^{-\frac{ikr^2}{2f}} J_0\left(\frac{kr\rho}{z}\right) r dr, \quad (S1)$$

where $R$ is the radius of the focusing lens, $u_0(r)$ is the electric field incident on the focusing lens, the exponential item $\exp(-ikr^2/f)$ is the equivalent phase factor of the low NA lens which is located at $z=0$, which implies that the electric field at the focal plane is obtained by setting $z=f$ in Eq. (S1). According to the concept of anti-resolution proposed in the main text, we should manipulate its PSF to realize the anti-resolution at the focal plane so that the electric field at the focal plane should be paid much attention. In Eq. (S1) by setting $z=f$ and omitting its constant factor $i2\pi/(\lambda f) \cdot \exp[ik(z+0.5\rho^2/f)]$, we have the electric field at the focal plane

$$\begin{aligned}
U(\rho) &= \int_0^R u_0(r) J_0\left(\frac{kr\rho}{f}\right) r dr = \sum_{n=1}^N \int_{r_{n-1}}^{r_n} (-1)^n J_0\left(\frac{kr\rho}{f}\right) r dr \\
&= \sum_{n=1}^N (-1)^n \left(\frac{f}{k\rho}\right)^2 [tJ_1(t)]_{k\rho r_{n-1}/f}^{k\rho r_n/f}, \quad t = k\rho r/f \\
&= \sum_{n=1}^N (-1)^n \left[ r_n^2 \frac{J_1(k\rho r_n/f)}{k\rho r_n/f} - r_{n-1}^2 \frac{J_1(k\rho r_{n-1}/f)}{k\rho r_{n-1}/f} \right] \\
&= f^2 \sum_{n=1}^N (-1)^n \left[ \sin^2\theta_n \frac{J_1(k\rho \sin\theta_n)}{k\rho \sin\theta_n} - \sin^2\theta_{n-1} \frac{J_1(k\rho \sin\theta_{n-1})}{k\rho \sin\theta_{n-1}} \right], \quad \sin\theta_n = r_n/f \\
&= f^2 \left[ (-1)^N \sin^2\theta_N \frac{J_1(k\rho \sin\theta_N)}{k\rho \sin\theta_N} + 2\sum_{n=1}^{N-1} (-1)^n \sin^2\theta_n \frac{J_1(k\rho \sin\theta_n)}{k\rho \sin\theta_n} \right]
\end{aligned} \quad (S2)$$

where the incident beam is uniform and modulated by a $N$-belt binary phase so that $u_0(r)=\exp(i\varphi(r))$, $\sin\theta_n=r_n/f$ for the focusing lens obeys the sine condition, $\theta_n$ is the focusing angle, with $\theta_0=0$ and the maximum angle $\theta_N=\sin^{-1}(NA)$, as shown by Eq. 5 in the main text. Interestingly, we find that Eq. (S2) provides an analytical model, without any integral involved, to approximate the electric field at the focal plane. By using Eq. (S2), we can obtain the arbitrary intensity pattern when changing the unknown parameter $\theta_n$ ($n=1, 2, …, N-1$).

Before we introduce the method to obtain the anti-resolution at the focal plane, the physical concept of anti-resolution should be revisited on basis of the Eq. (S2). In anti-resolution, its PSF as shown in Fig. 1d has the zero intensity at $\rho=0$ (suppressing the mainlobe intensity), the approaching zero intensity in the range $0<\rho<D/2$ (widening the mainlobe) and the largest sidelobe intensity (enhancing the sidelobe). In order to realize the goal of suppressing the mainlobe intensity in anti-resolution, we just set the intensity to be zero at $\rho=0$, leading to the equation

$$(-1)^N \sin^2\theta_N + 2\sum_{n=1}^{N-1}(-1)^n \sin^2\theta_n = 0. \quad (S3)$$

In Eq. (S3), the $N$-1 unknown parameters $\theta_n$ ($n$=1, 2, …, $N$-1) indicate the infinite solutions to realize the goal of suppressing the mainlobe, which implies that it is possible to fix one solution if more constraints is imposed to Eq. (S3), e.g. reserving $N$-2 zero-intensity locations in the region 0<$\rho$<$D$/2 so as to widen the mainlobe (the second goal in anti-resolution). For the third goal of enhancing the sidelobe, we do not need to make any measure to realize it because the high sidelobe is the natural result in the viewpoint of energy conservation for the approaching-zero intensity in the range 0≤$\rho$<$D$/2. Therefore, it is physically feasible to construct a PSF with anti-resolution. In fact, the zero intensity in the region 0≤$\rho$<$D$/2 shown in Fig. (S3) is mainly attributed to the local destructive interference caused by the 0-π phase modulation of binary element.

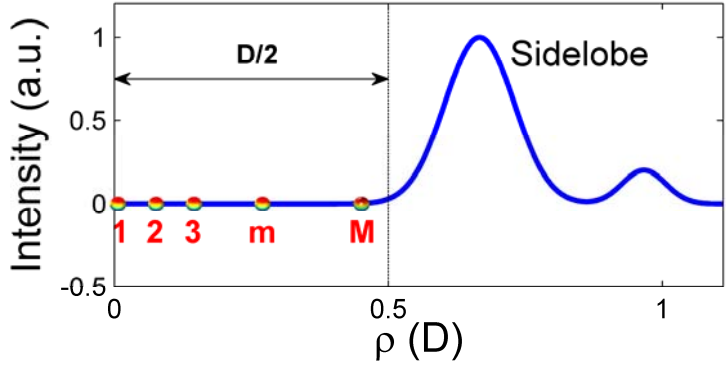

Fig. S3 The radial (along $\rho$) pattern of the anti-resolution PSF. The prescribed positions from 1 to **M** (**M**=$N$-1) in the range 0≤$\rho$<$D$/2 are shown by the color spots and located at the zero-intensity part of the intensity line (blue).

In order to show the design conveniently, we simplify Eq. (S2) by some characteristic functions:

$$A(\rho) = (-1)^N \sin^2\theta_N \frac{J_1(k\rho\sin\theta_N)}{k\rho\sin\theta_N}, \quad (S4)$$

$$C(\theta_n) = 2\cdot(-1)^n \sin^2\theta_n, \quad (S5)$$

$$S(\rho,\theta_n) = \frac{J_1(k\rho\sin\theta_n)}{k\rho\sin\theta_n}, \quad (S6)$$

so that we have

$$U(\rho) = A(\rho) + \sum_{n=1}^{N-1} C(\theta_n)S(\rho,\theta_n), \quad (S7)$$

without the constant item $f^2$ in Eq. (S2). According to the above discussion, beyond the central position $\rho$=0, we still need to prescribe $N$-2 zero-intensity positions in the region $0<\rho<D/2$, which are shown from 2 to M in Fig. S3. The prescribed zero-intensity positions are labeled as $m$ ($m$=1, 2, …, M) to distinguish from the belt label $n$, resulting that the prescribed zero-intensity positions can be labeled as $\rho_m$ ($m$=1, 2, …, M). According to Eq. (S7), we have the M equations for M positions as follows:

$$A(\rho_1)+C(\theta_1)S(\rho_1,\theta_1)+C(\theta_2)S(\rho_1,\theta_2)+\cdots+C(\theta_n)S(\rho_1,\theta_n)+\cdots+C(\theta_{N-1})S(\rho_1,\theta_{N-1})=U(\rho_1)=0$$
$$A(\rho_2)+C(\theta_1)S(\rho_2,\theta_1)+C(\theta_2)S(\rho_2,\theta_2)+\cdots+C(\theta_n)S(\rho_2,\theta_n)+\cdots+C(\theta_{N-1})S(\rho_2,\theta_{N-1})=U(\rho_2)=0$$
$$\vdots$$
$$A(\rho_m)+C(\theta_1)S(\rho_m,\theta_1)+C(\theta_2)S(\rho_m,\theta_2)+\cdots+C(\theta_n)S(\rho_m,\theta_n)+\cdots+C(\theta_{N-1})S(\rho_m,\theta_{N-1})=U(\rho_m)=0$$
$$\vdots$$
$$A(\rho_M)+C(\theta_1)S(\rho_M,\theta_1)+C(\theta_2)S(\rho_M,\theta_2)+\cdots+C(\theta_n)S(\rho_M,\theta_n)+\cdots+C(\theta_{N-1})S(\rho_M,\theta_{N-1})=U(\rho_M)=0$$

which can be simplified further by a matrix equation

$$\mathbf{SC} = \mathbf{A}, \qquad (S8)$$

where $\mathbf{S}$ is an M×(N-1) matrix with its matrix element $S_{mn}=S(\rho_m,\theta_n)$, $\mathbf{C}$ is a (N-1)×1 matrix with its matrix element $C_n=C(\theta_n)$ and $\mathbf{A}$ is an M×1 matrix with its element $A_n=-A(\rho_m)$. Eq. (S8) is a non-linear matrix equation because the $\mathbf{S}$ and $\mathbf{C}$ are dependent on the unknown parameters $\theta_n$. In Eq. (S8), the number of unknown parameter and equations are the same so that it has the only solution. For a special position $\boldsymbol{\rho}=[\rho_1, \rho_2, …, \rho_M]$, which is up to the customized requirement in the size of D shown in Fig. S3, we just solve the non-linear equation shown by Eq. (S8) to finish the design of binary phase by fixing the angles $\theta_n$ ($n$=1, 2, …, N-1). Comparing with the solution of a linear matrix equation, the non-linear matrix equation in Eq. (S8) can not be solved by the simple matrix-inversion technique that is used to solve the linear matrix equation. In fact, for the non-linear matrix equation, it generally has no analytical solution but its numerical solution can be obtained in a quite easy way by using the well-developed Newton theory, which is widely used in the relative engineering problems [3]. For the detailed process to solve Eq. (S8), we ignore it here because there are so many references about it. In addition, one can also find the special packages to solve the non-linear matrix equation in some commercial computing software, e.g. MATLAB or MATHEMATICA. For the detailed information about our codes used to solve Eq. (8), one can refer to our new paper [4]. Therefore, the solution of Eq. (S8) is not a troublesome issue. Sometimes, for one special position $\boldsymbol{\rho}$ (e.g. $\rho_m$ and $\rho_{m+1}$ are too close), the mathematical solution, provided by Newton's theory, of Eq. (S8) might be not a physical solution and we should drop it automatically. For a physical solution, it should satisfy the condition: $0<\sin\theta_n<\sin\theta_N$.

In Fig. 5a, we show the four sets of designed binary-phase plates for generating the anti-resolution PSF using the problem in Eq. (S8). We choose the zero-intensity position at $\rho$=[0, 22.5μm, 45μm, 67.5μm, 90μm] for *No*. 1, $\rho$=[0, 27.5μm, 75μm, 82.5μm, 110μm] for *No*. 2, $\rho$=[0, 32.5μm, 65μm, 97.5μm, 130μm] for *No*. 3 and $\rho$=[0, 37.5μm, 75μm, 112.5μm, 150μm] for *No*. 4. The NA of focusing lens is fixed to 0.005 when we design the four sets of binary-phase plates. The data for the four sets of binary-phase plates are shown in Fig. 5a. For the design, we have to claim several points:

1) Both number of equations and unknown parameters in Eq. (S8) are the same (N-1) so that it has the only solution, but in real design by solving the Eq. (S8), the number of equations can be larger than (N-1) by prescribing more zero-intensity position in $\rho$ (as shown in the above examples: four unknown parameters $\theta_n$ (*n*=1, 2, 3 and 4) but we prescribe five zero-intensity positions). The reason for this is to suppress the null field in the range of D. Another issue caused by this is that Eq. (S8) has no exact solution for the more zero-intensity position at $\rho$. But, this will not lead to the failure of designing the anti-resolution PSF because we only need the approximate solution numerically with the condition that the null-field ($0 \leq \rho < D$) intensity is much below the sidelobe intensity. In our case, we take it as the null field when its intensity is below $10^{-3}$ of the sidelobe intensity. Therefore, although we can not get the exact solution when we increase the number of the prescribed zero-intensity position, we can still get another solution to realize the anti-resolution PSF with more low intensity in the null-field region.

2) Because the binary-phase plate has the intrinsic property that can generate the null field at its Fraunhofer diffraction region, the focusing lens can not change the property and just transfer the Fraunhofer diffraction region to the focal region of focusing lens. Therefore, we design the four sets of binary-phase plates by using the focusing lens with NA=0.005, but we show the axial size ($d_0$) of the null field in the focusing lens with NA=0.0075 in order to match the experimental focusing lens in Fig. 4.

Finally, we have to emphasize that the method to design the binary phase plate for realizing the anti-resolution PSF is optimization-free because we can solve the inverse problem described in Eq. (S8) numerically by the well-developed Newton theory. It is worthy to point out that the optimization-free design method to control the optical field by a binary plate has been introduced to construct a super-oscillatory lens for realizing the super-resolution focusing.

**5. Generation of optical capsule based on the anti-resolution PSF**

In the last section, we have introduced the optimization-free design of binary phase for realizing the anti-resolution PSF. Now, we explain the physical reason for generation of optical capsule when the anti-resolution PSF is achieved at the focal plane.

First, we revisit the different focusing behavior along transverse and axial direction for a lens. Without loss of generality, we analyze the intensity at focal plane $z=f$ for the transverse direction and the on-axis intensity with $r=0$ for the axial direction. For the uniform illumination without any phase or amplitude modulation by binary elements, we can get the analytical form of the intensity at $z=f$ and $r=0$ according to Eq. (S1).

$$I(\rho, z=f) = \left| \int_0^R J_0\left(\frac{kr\rho}{f}\right) r dr \right|^2 = R^4 \left[ \frac{J_1(k\rho R/f)}{k\rho R/f} \right]^2, \quad (S9)$$

$$I(\rho=0, z) = \left| \int_0^R e^{\frac{ikr^2}{2z}} e^{-\frac{ikr^2}{2f}} r dr \right|^2 = \frac{R^4}{4} \left\{ \frac{\sin\left[R^2 k \left(\frac{1}{z}-\frac{1}{f}\right)\right]}{R^2 k \left(\frac{1}{z}-\frac{1}{f}\right)} \right\}^2, \quad (S10)$$

where $I(\rho,z=f)$ and $I(\rho=0,z)$ show the intensity profiles at focal plane and on axis with $\rho=0$, respectively. From Eqs. (S9) and (S10), we can derivate the size (from the focal point to the first zero-intensity) in radial and axial directions. For the radial size at the focal plane, the first zero-intensity point is located at $k\rho R/f=3.84$, which implies the radial size is

$$\Delta \rho = \frac{0.61\lambda}{NA}, \quad (S11)$$

where $NA \approx R/f$ for a low NA lens as shown by Fig. S4a- S4b. For the axial size with $\rho=0$, the first zero-intensity point is located at $R^2 k(1/z-1/f)=\pi$, which implies that the axial size (Fig. S4a-S4c) is

$$\Delta z = \left| \frac{\lambda}{\lambda/f + 2(R/f)^2} \right| \approx \frac{0.5\lambda}{NA^2}, \quad (S12)$$

From the radial and axial size denoted by Eqs. (S11) and (S12), one can see that the radial size is proportional to $1/NA$ while the axis size is proportional to $1/NA^2$. Because the lens' NA is smaller than 1, the spot in radial direction is smaller than that in axial direction as shown in Fig. (S4), which indicates that the focusing lens has a tighter confinement in radial direction than axial direction. It is a very important as well as fundamental conclusion in optical focusing of lens. Therefore, for the well-built anti-resolution PSF at the focal plane, just standing on this physical conclusion without further investigation into the intensity distribution in axis region, we can have an intuitionistic prediction: the zero-intensity region in axis direction with $\rho=0$ must be much larger than that in its radial region.

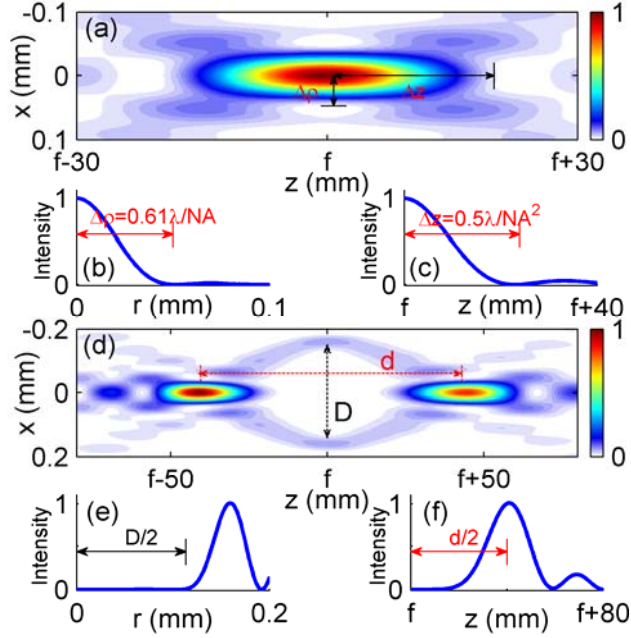

Fig. S4 The focusing properties of a lens. (a) The intensity profile at the x-z plane of focusing field by a lens. (b) The radial intensity profile of the focused spot at the focal plane in (a). (c) The axial intensity profile at $\rho=0$ in (a). (d) The intensity profile of optical capsule in the focal region by using a lens and a binary-phase plate. D and d is the radial and axial sizes (from one hotspot to another) of the optical capsule. (e) The radial intensity profile of the optical capsule at the focal plane in (d). (f) The axial intensity profile of the optical capsule at $\rho=0$ in (d).

Next, we check the above prediction through the analytical simulation by using Eq. (S1) because this theoretical prediction proves the generation of optical capsule on the basis of anti-resolution PSF. In fact, the simulation results in Fig. 5c-5c have proved the prediction that the transversal (D) and axial sizes ($d_0$) are proportional to $1/NA$ and $1/NA^2$, respectively. Because NA<1, the axial size ($d_0$) is always larger than the transversal size (D). Although we can find the undoubtable proof to verify the existence of the axial null-field from the simulation in Fig. 5c, we can also find another physical proof based on the uniform variation of energy flux of light in the homogenous medium. In section 4, we showed the generation of the anti-resolution PSF with the null-field region $0\leq\rho<D$ at the focal plane. If we move the viewing position to one out-of-plane ($z\neq f$) position, we can predict the existence of null field at this out-of-plane position because of the uniform variation of energy flux of light in air as shown in Fig. S4d. The null-field region will exist until the out-of-plane position is located at the two ends of the optical capsule. Because the axial confinement of focusing lens is much weaker than its transversal confinement, as shown by Eqs. S11 and S12, the null field in axial direction has the longer extension than that in transversal direction. Therefore, an optical capsule with long boundary in axial direction and short boundary in transversal direction is formed after an anti-resolution PSF has been generated at the focal plane.

The sections 4 and 5 are contributed to explain the physical reason for the generation of optical capsule by using the novel anti-resolution PSF concept in the optical imaging system with focusing lens, which motivates us to propose the optical capsule in theory and verify it in experiment. However, the basic reason is the substantial property of the binary-phase plate in generating the optical capsule at its Fraunhofer-diffraction region as depicted in the main text.